\documentclass[preprintnumbers,showpacs, showkeys, twocolumn,secnumarabic,amssymb,amsmath,amsfonts,aps, prl]{revtex4-1}

\usepackage{graphicx}

\usepackage{amsmath}
\usepackage{amsfonts}
\usepackage{amssymb}
\begin{document}

\newtheorem{theorem}{Theorem}
\newtheorem{remark}{Remark}
\newcommand{\non}{\nonumber}
\newcommand{\be}{\begin{equation}}
\newcommand{\ee}{\end{equation}}
\newcommand{\bq}{\begin{eqnarray}}
\newcommand{\eq}{\end{eqnarray}}
\newcommand{\lps}{\langle}
\newcommand{\rps}{\rangle}
\newcommand{\vb}{\bar{v}}
\newcommand{\D}{\mathrm{d}}

\everymath{\displaystyle}

\title{On the apparent failure of the topological theory of phase transitions}

\date{\today}
\author{Matteo Gori}
\email{gori6matteo@gmail.com}
\affiliation{Aix-Marseille University, CNRS Centre de Physique Th\'eorique UMR 7332,
Campus de Luminy, Case 907, 13288 Marseille Cedex 09, France}

\author{Roberto Franzosi}
\email{roberto.franzosi@ino.it}
\affiliation{Qstar, Istituto Nazionale di Ottica, largo E. Fermi 6, 50125 Firenze, Italy}

\author{Marco Pettini}
\email{pettini@cpt.univ-mrs.fr}
\affiliation{Aix-Marseille University, CNRS Centre de Physique Th\'eorique UMR 7332,
Campus de Luminy, Case 907, 13288 Marseille Cedex 09, France}

\begin{abstract}
The topological theory of phase transitions has its strong point in two theorems proving that,
for a wide class of physical systems, phase transitions necessarily stem from topological changes of
some submanifolds of configuration space. It has been recently argued that the $2D$ lattice $\phi^4$-model  provides a counterexample that falsifies this theory. It is here shown that this is not the case: the phase transition of this model stems from an asymptotic  ($N\to\infty$) change of topology of the energy level sets, in spite of the absence of critical points of the potential in correspondence of the transition energy.  

\end{abstract}
\maketitle
\begin{figure}[h!]
 \centering
 \includegraphics[scale=0.35,keepaspectratio=true]{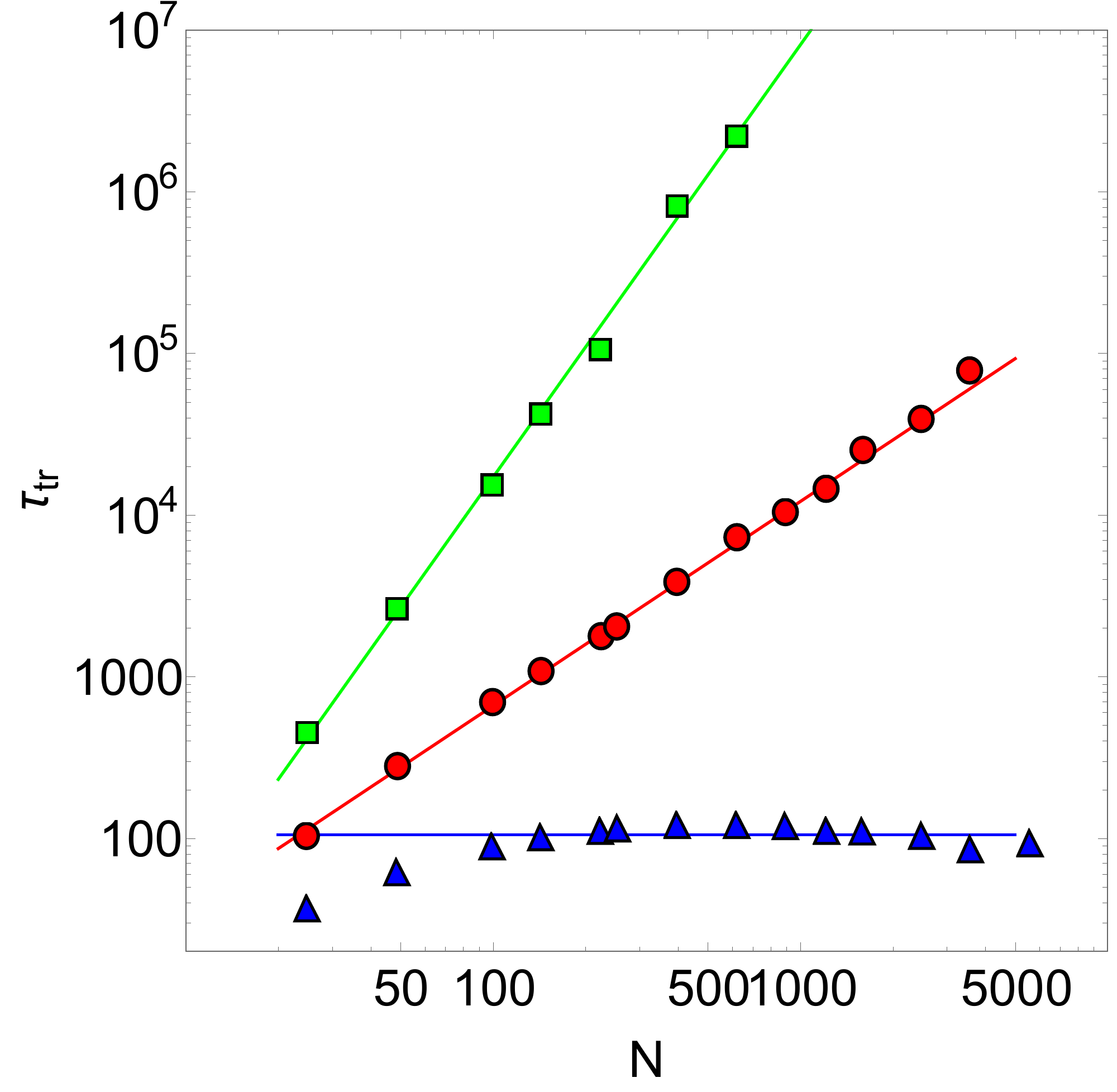}
 \caption{(Color online) Average trapping time $\tau_{tr}$ of the magnetization vs. the number of lattice sites $N$ for the 2D $\phi^4$-model. Different data series refer to different values of the energy per degree of freedom $\varepsilon$: $\varepsilon=8$ (squares), $\varepsilon=10$ (circles), both below the transition energy $\varepsilon_c=11.1$, and $\varepsilon=12$ (triangles), above the transition energy.}
\label{taueffe}
\end{figure}

\noindent{\bf Introduction. } Hamiltonian flows ($H$-flows) can be seen as geodesic flows on suitable Riemannian manifolds \cite{PettiniBook}.  Within this framework, it turns out that: \textit{i)} the Riemannian geometrization of $H$-flows allows to explain the origin of Hamiltonian chaos and, in some cases, also makes it possible the analytic computation of the largest Lyapounov exponent (LLE); \textit{ii)} the energy dependence of the LLE displays peculiar patterns for those systems which undergo a thermodynamic phase transition (TDPT) .
Hence the question naturally arises of whether and how these manifolds "encode" the fact that their geodesic flows/$H$-flows are associated or not with a TDPT. By following this conceptual pathway, eventually one is led to take into account the topological properties of certain submanifolds of phase space. The existence of a relationship between thermodynamics and phase space - or configuration space - topology is provided by the following exact formula 
\[
S_N(v) =({k_B}/{N}) \log \left[ \int_{M_v}\ d^Nq\right] 
\]
\[
=\frac{k_B}{N} \log \left[ vol
[{M_v \setminus\bigcup_{i=1}^{{\cal N}(v)} \Gamma(x^{(i)}_c)}]\ +
\sum_{i=0}^N w_i\ \mu_i (M_v)+ {\cal R} \right]  ,\label{exactS}
\]
where $S$ is the configurational entropy, $v$ is the potential energy per degree of freedom,  and the
$\mu_i(M_v)$ are the Morse indexes (in one-to-one correspondence
with topology changes) of the submanifolds $\{ M_{v}=V_N^{-1}((-\infty,v])\}_{v \in{\Bbb R}}$ 
of configuration space; in square brackets: the first term is the result of the excision of certain neighborhoods of the critical points of the interaction potential from  $M_{v}$; the second term
is a weighed sum of the Morse indexes, and the third term is a smooth function of $N$ and $v$.
It is evident that sharp changes in the potential energy pattern of at least some of
the $\mu_i(M_v )$ (thus of the way topology changes with $v$) affect $S(v)$ and its
derivatives, hence some \textit{sufficient} condition to entail a TDPT can be obtained. Then one can wonder if topological changes are also \textit{necessary} for the appearance of a TDPT. A few exactly solvable models \cite{PettiniBook} and two theorems \cite{prl1,TH1,TH2} answered in the affirmative: topological changes of the $M_v$ (or equivalently of the $\Sigma_v$) are \textit{necessary but not sufficient} conditions to break the uniform convergence of Helmholtz free energy, and thus to entail a TDPT.
The practical interest of the topological theory of PTs could nowadays turn from potential to actual thanks to recent developments of powerful computational methods in algebraic topology, like those of \textit{persistent homology} \cite{Carlsson, Chazals}.\\
However, it has been recently argued \cite{kastner} against this theory on the basis of the observation that the second order phase transition of the $2D$ lattice $\phi^4$-model occurs at a critical value $v_c$ of the potential energy density which belongs to a broad interval of $v$-values void of critical points of the potential function. In other words, the $\{ \Sigma_{v<v_c}\}_{v \in{\Bbb R}}$ are diffeomorphic to the  $\{ \Sigma_{v>v_c}\}_{v \in{\Bbb R}}$ so that no topological change seems to correspond to the phase transition.
In spite of the claim that this counterexample falsifies the theory, in the present paper we discuss how a suitable refinement can fix the problem paving the way to a more general formulation of the theory itself. 
Let us remark that a counterexample to a theory does not necessarily mean that it has to be discarded, to the contrary, a counterexample can stimulate a refinement of a theory.
An instance, which is not out of place in the present context, is the famous counterexample that J.Milnor gave against De Rham's cohomology theory (the two manifolds $M = \mathbb{S}^2\times\mathbb{S}^4$, product of spheres, and $N = \mathbb{C}P^3$, complex-projective space, are neither diffeomorphic nor homeomorphic yet have the same cohomology groups). The introduction of the so called "cup product" fixed the problem and saved the theory making it more powerful. 
Let us remark that the two basic theorems in Refs.\cite{prl1,TH1,TH2} rely on the assumption of diffeomorphicity at any arbitrary \textit{finite} $N\in\mathbb{N}$ of any pair of $\Sigma_{v,N}$, with $v\in [v_0,v_1]$, but no assumption is made about the asymptotic ($N\to\infty$) properties of the diffeomorphicity relation among the $\Sigma_{v,N}$. Thus we proceed to numerically investigate on this aspect.\\
\noindent{\bf The lattice $\phi^4$ model. } 
The model of interest, considered in Refs.\cite{kastner,CCP},  is defined by the Hamiltonian 
\begin{equation}
{\cal H} ( p, q )= \sum_{\bf i}   \frac{p_{\bf i}^2}{2}  + V(q)
\label{Hphi_2}
\end{equation} 
where the potential function $V(q)$ is
\begin{equation}
V(q)=\sum_{{\bf i}\in{\Bbb Z}^D}\left( - \frac{\mu^2}{2} q_{\bf i}^2 +
\frac{\lambda}{4!} q_{\bf i}^4 \right) + \sum_{\langle {\bf 
ik}\rangle\in{\Bbb Z}^D}\frac{1}{2}J (q_{\bf i}-q_{\bf k})^2\ ,
\label{potfi4}
\end{equation}
with $\langle {\bf ik}\rangle$ standing for nearest-neighbor sites on a $D$ dimensional lattice. This
system has a discrete ${\Bbb Z}_2$-symmetry and short-range
interactions; therefore, according to the Mermin--Wagner theorem,
in $D=1$ there is no phase transition whereas in $D=2$ there is a a second order 
symmetry-breaking transition,  with nonzero critical temperature, of the same
universality class of the 2D Ising model. \\
The numerical integration of the equations of motion derived from Eqs.\eqref{Hphi_2} and \eqref{potfi4} has been performed for $D=2$, with periodic boundary conditions, using a bilateral symplectic integration scheme \cite{Lapo} with time steps chosen so as to keep energy conservation within a relative precision of $\Delta E/E\simeq 10^{-6}$. The model parameters have been chosen as follows: $J=1$, $\mu^2 = 2$, and $\lambda = 3/5$. By means of standard computations as in Refs.\cite{CCCPPG} and \cite{CCP}, and for the chosen values of the parameters, the $2D$ system undergoes the symmetry-breaking phase transition at a critical energy density value  $\varepsilon_c=E_c/N\simeq 11.1$, correspondingly the critical potential energy density value is $v_c=\langle V\rangle_c/N\simeq 2.2$ \cite{nota}.  Random initial conditions have been chosen. With respect to the already performed numerical simulations we have here followed the time evolution of the order parameter ("magnetization")
\begin{equation}
{M}=\frac{1}{N}\sum_{\bf i} q_{\bf i}\ .
\end{equation}
This vanishes in the symmetric phase, that is for $\varepsilon > \varepsilon_c$, whereas it takes a positive or negative value in the broken symmetry phase, that is for $\varepsilon < \varepsilon_c$. However, at finite $N$ the order parameter can flip from positive to negative and viceversa. This flipping is associated with a trapping phenomenon of the phase space trajectories alternatively in one of the two subsets of the constant energy surfaces which correspond to positive and negative magnetization, respectively. This phenomenon has been investigated by computing the average trapping time $\tau_{tr}$  for different lattice sizes, and choosing values of $\varepsilon$ just below and just above $\varepsilon_c$ . The results are displayed in Figure \ref{taueffe}. Denote with $\varphi^H_t:\Sigma_E\to\Sigma_E$ the ${\cal H}$-flow, with $\Sigma_E={\cal H}^{-1}(E)$ a constant energy hypersurface of phase space,  with ${\cal M}_E^+\subset\Sigma_E$ the set of all the phase space points for which $M\ge\eta >0$,  with ${\cal M}_E^-\subset\Sigma_E$ the set of all the phase space points for which $M\le -\eta <0$, and with ${\cal M}_E^\eta\subset\Sigma_E$ a transition region, that is, the set of all the phase space points for which $-\eta \le M\le\eta$, with $\eta\ll \langle\vert M\vert\rangle$ \cite{emmemu}.  Thus $\Sigma_E ={\cal M}_E^+ \cup  {\cal M}_E^-\cup{\cal M}_E^\eta$. From the very regular functional dependences of $\tau_{tr}(N)$ reported in Figure \ref{taueffe}, we can see that:\\
\textit{At $\varepsilon < \varepsilon_c$, for any given $\tau_{tr} >0$ there exists an $N(\tau_{tr})$ such that for any $N>N(\tau_{tr})$ and $t\in [0,\tau_{tr}]$ we have   $\varphi^H_t({\cal M})_E^\pm = {\cal M}_E^\pm$ }. \\ In other words, below the transition energy density the  subsets ${\cal M}_E^\pm$ of the constant energy surfaces $\Sigma_E$ appear to be \textit{invariant} for the ${\cal H}$-flow on a finite time scale $\tau_{tr}$, with the remarkable fact that $\tau_{tr}\to\infty$ in the limit $N\to\infty$ \cite{nota3}. Formally this reads as 
\begin{eqnarray}
\forall A\subset{\cal M}_E^+, \forall B\subset{\cal M}_E^- &{\rm \; and\;}& t\in [0,\tau_{tr}(N)]\nonumber \\
{\rm it\, is\,} \ \   \varphi^H_t(A)\cap B &=& \emptyset \ .
\label{TT1}
\end{eqnarray} 
To the contrary:\\
\textit{At $\varepsilon > \varepsilon_c$, there exists a $\tau_{tr}^0 >0$ such that for any $N$ and }
\begin{eqnarray}
\forall A\subset{\cal M}_E^+, \forall B\subset{\cal M}_E^- &{\rm \; and\;}& t>\tau_{tr}^0 \nonumber \\
{\rm it\, is\,} \ \   \varphi^H_t(A)\cap B &\neq & \emptyset \ .
\label{TT2}
\end{eqnarray} 

Since $\Sigma_E ={\cal M}_E^+ \cup  {\cal M}_E^-\cup{\cal M}_E^\eta$, and since the residence times in the transition region are found to be very short and independent of $N$ - so that the relative measure ${\sl meas}({\cal M}_E^\eta)/{\sl meas}({\cal M}_E^\pm)$ vanishes in the limit $N\to\infty$ - Eq. \eqref{TT1} means that below the transition energy the \textit{topological transitivity} of $\Sigma_E$ is broken up to a time $\tau_{tr}(N)$ --  which is divergent with $N$. To the contrary, above the transition energy the $\Sigma_E$ are \textit{topologically transitive} \cite{toptran}.  The asymptotic breaking of topological transitivity at $\varepsilon < \varepsilon_c$, that is the divergence of $\tau_{tr}(N)$ in the limit $N\to\infty$,  goes together with asymptotic ergodicity breaking due to the ${\Bbb Z}_2$-symmetry breaking. Moreover, on metric and compact topological spaces,  topological transitivity is equivalent to \textit{connectedness} of the space \cite{toptran}, so the loss of topological transitivity entails the loss of connectedness, that is, a major topological change of the space. And if we denote by $H_{\tau}^0(\Sigma^N_E;{\Bbb R})$ the "finite time zeroth cohomology space" of $\Sigma_E$, for $\tau < \tau_{tr}(N)$ we have $b_0=\dim H_{\tau}^0(\Sigma^N_E;{\Bbb R}) = 2$ at $\varepsilon < \varepsilon_c$, and $b_0=\dim H_{\tau}^0(\Sigma^N_E;{\Bbb R}) = 1$ at $\varepsilon > \varepsilon_c$. The dimension of this cohomology space (the Betti number $b_0$) counts the number of connected components of $\Sigma_E$ and is invariant under diffeomorphisms of the $\Sigma_E$. Hence the asymptotic jump of a diffeomorphism invariant across the phase transition point, which can be deduced by our numerical computations, means that the $\Sigma_E$ undergo an \textit{asymptotic loss of diffeomorphicity}, in the absence of critical points \cite{nota2} of the potential $V(q)$. Now, the breaking of topological transitivity of the $\Sigma_E$ implies the same phenomenon for configuration space and its submanifolds $\Sigma_v=V^{-1}(v)$ (potential level sets). These level sets are the basic objects, foliating configuration space, that enter the theorems in \cite{prl1,TH1,TH2}, and represent the nontrivial topological part of phase space. The link of these geometric objects with microcanonical entropy is given by 
\begin{equation}
S(E) =\frac{k_B}{2N} \log \int_0^E d\eta \int d^Np\  \delta (\sum_{\bf i} p_{\bf i}^2/2 - \eta ) \int_{\Sigma_{E-\eta}} \frac{d\sigma}{\Vert\nabla V\Vert} \ .
\end{equation}
As $N$ increases the microscopic configurations giving a relevant contribution to the entropy, and to any  microcanonical average, concentrate closer and closer on the level set $\Sigma_{\langle E-\eta\rangle}$. Therefore, it is interesting to make a direct numerical analysis on these level sets at different $N$ values to find out - with a purely geometric glance - how configuration space asymptotically breaks into two disjoint components. The intuitive picture is that, approaching from above ($\varepsilon > \varepsilon_c$) the transition point,  some subset of each $\Sigma_v$ - a "high dimensional neck" related with ${\cal M}_E^\eta$ - should be formed which bridges the two regions ${\cal M}_v^+$ and ${\cal M}_v^-$. And this neck should increasingly shrink with increasing $N$. To perform this analysis we resort to a Monte Carlo algorithm constrained on any given $\Sigma_v$. This is obtained by  generating a Markov Chain with a Metropolis importance sampling of the weight $\chi=1/{\Vert\nabla V\Vert}$.   In order to check the validity of the intuitive idea of a neck which shrinks at increasing $N$, we have to identify some useful geometric quantities to be numerically computed. To do this we proceed as follows. Let us note that, in the absence of critical points in an interval $[a,b]$, the explicit form of the diffeomorphism $\boldsymbol{\xi}$ that maps one to the other the level sets $\Sigma_c=f^{-1}(c)$, $c\in[a,b]$, of a function $f:{\Bbb R}^N\to{\Bbb R}$ is explicitly given by \cite{hirsch}
\begin{equation}
\frac{dx^i}{dc} = \xi^i(x) =\frac{\nabla^i f(x)}{\Vert\nabla f(x)\Vert^2}\ ,
\label{hirschflow}
\end{equation}
and this applies as well to the energy level sets in phase space as to the potential level sets in configuration space.
If we consider an infinitesimal change of potential energy $v\rightarrow v+\epsilon_{v}$ with $\vert\epsilon_{v}\vert/v\ll 1$, and denote with $\delta (q)$ the field of local distances  between two level sets $\Sigma_v$ and $\Sigma_{v+\epsilon_{v}}$, from 
$q^i(v+\epsilon_{v})=q^i(v) + \xi^i \epsilon_{v}$ and using Eq.\eqref{hirschflow}, at first order in $\epsilon_{v}$,  we get $\delta(q)=\epsilon_{v}/\Vert \nabla V\Vert_q=\epsilon_{v}\chi(q)$. 
Moreover the divergence $\mathrm{div}\boldsymbol{\xi}$ in euclidean configuration space can be related with the variation rate of the measure of the 
microcanonical area  $\mathrm{d}\mu=\chi\mathrm{d}\sigma$ over regular level sets $\Sigma_{v}$.
The first variation formula  for the induced measure of the Riemannian area  $\mathrm{d}\sigma$ along the flow $q(v)$ reads \cite{LeeGeometry}:
\begin{equation}
\mathrm{d}\sigma(q(v+\epsilon_v))=\left(1-\epsilon_v-\chi M_1\right)\mathrm{d}\sigma(q(v))+o(\epsilon_v)
\end{equation}
where $M_1$ is the sum of the principal curvatures of $\Sigma_v$ that is given by
\begin{equation}
M_1=-\mathrm{div}\left(\dfrac{\nabla V}{\|\nabla V\|}\right).
\end{equation}
Applying the Leibniz rule, the first variation formula for the measure of the microcanonical area is
\begin{equation}
\begin{split}
&\mathrm{d}\mu(q(v+\epsilon_v))=\chi(q(v+\epsilon_v))\mathrm{d}\sigma(v+\epsilon_v)=\\
&=\left[1+\epsilon_v\left(-\chi M_1+\dfrac{(\nabla^i V)}{\|\nabla V\|}\nabla_i\chi\right)\right]\mathrm{d}\mu=\\
&=\left(1+\epsilon_v\mathrm{div}\boldsymbol{\xi}\right)\mathrm{d}\mu(q(v))
\end{split}
\end{equation}

\begin{figure}[h!]
 \centering
 \includegraphics[scale=0.35,keepaspectratio=true]{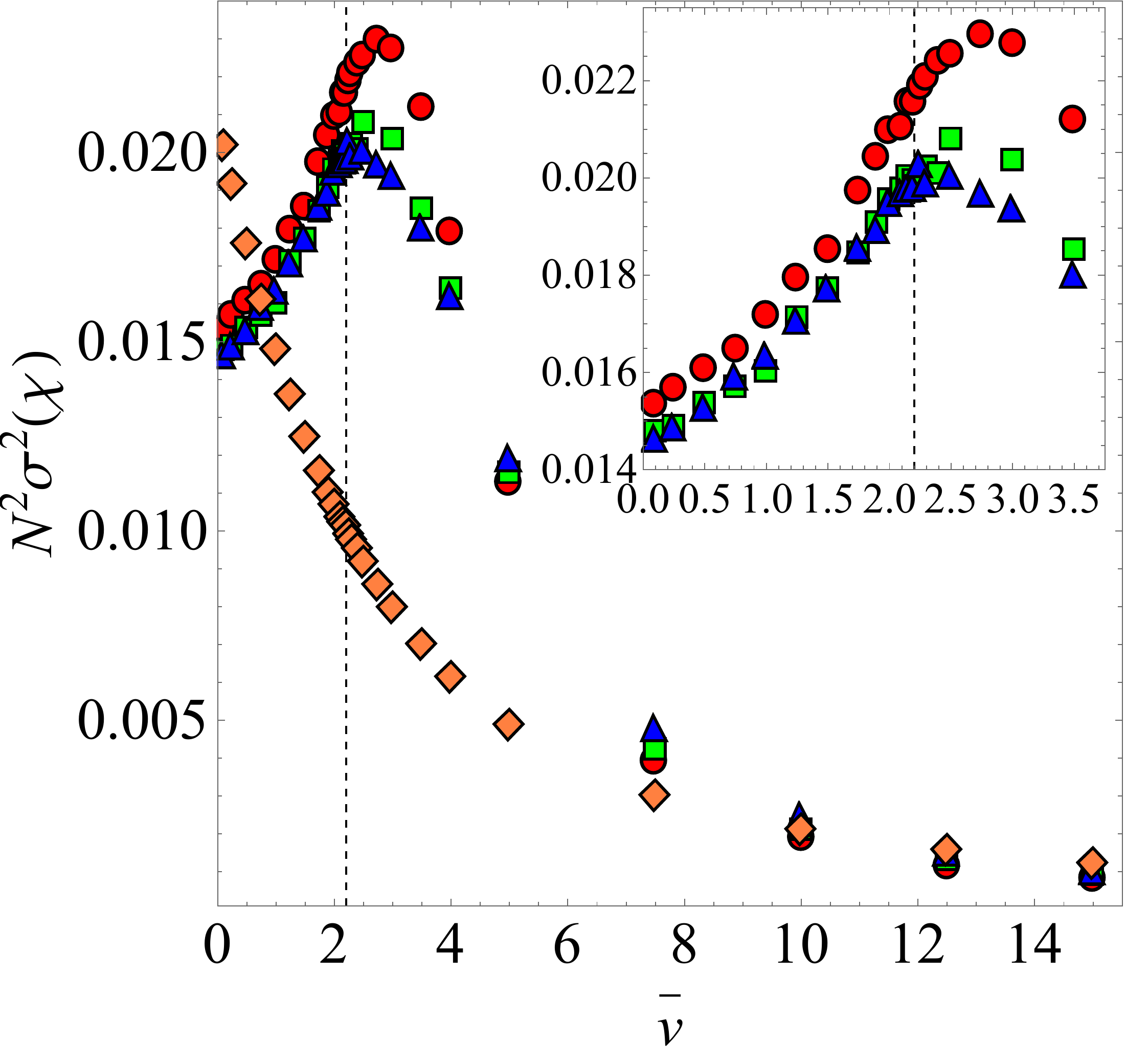}
\caption{(Color online) Variance of $\chi$ vs.  potential energy per degree of freedom $\bar{v}$ for 1D and 2D $\phi^4$-models, and for lattice sizes: $N=10\times 10$ (circles), $N=20\times 20$ (squares), $N=30\times 30$ (triangles) in the 2D case, and $N=900$ (rhombs) in the 1D case. The vertical dashed line indicates the phase transition point at $\bar{v}\simeq 2.2$.}
\label{var-chi}
\end{figure}

\begin{figure}[h!]
 \centering
 \includegraphics[scale=0.35,keepaspectratio=true]{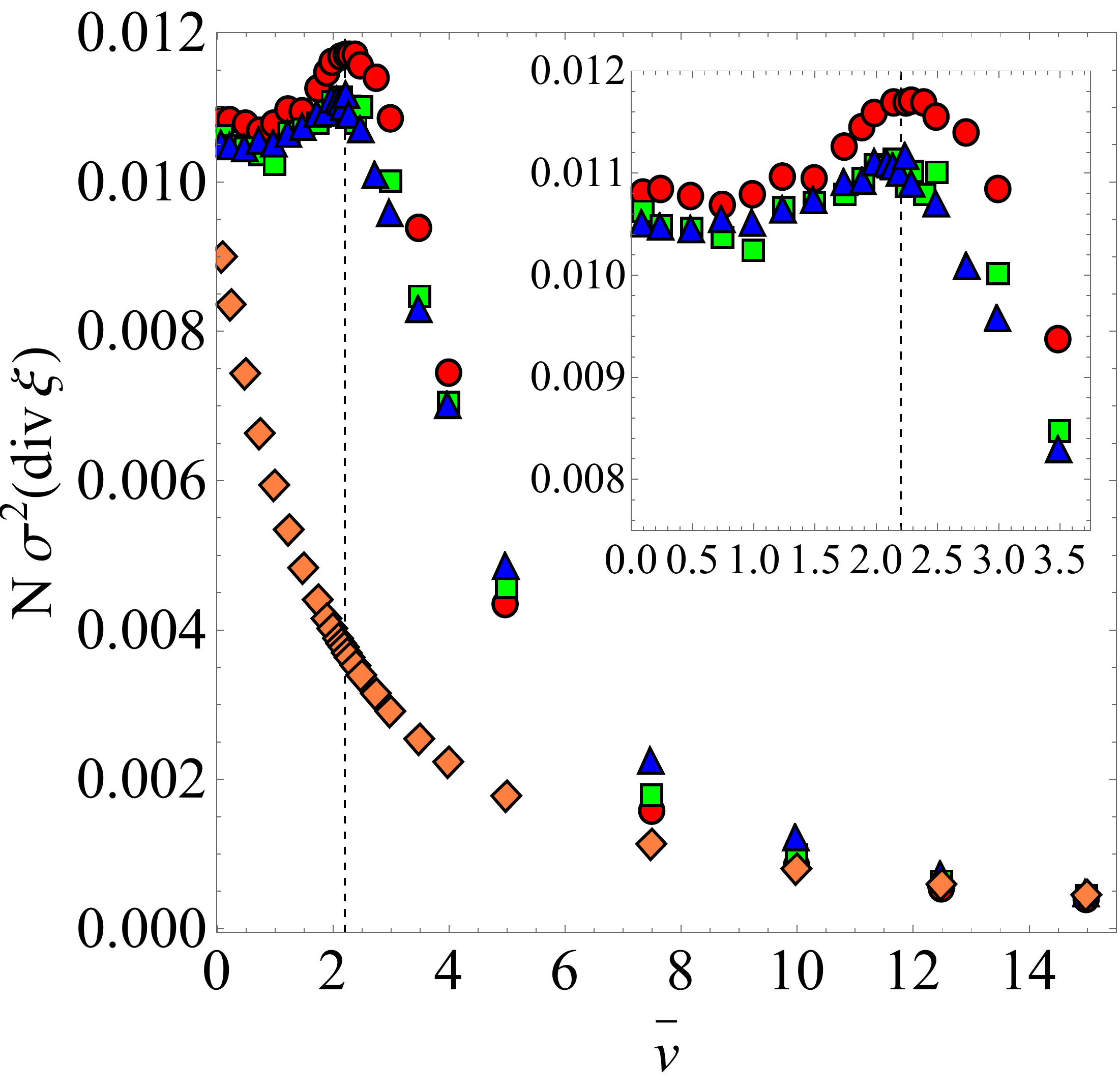}
\caption{(Color online) Variance of $\mathrm{div}\xi$ vs. potential energy per degree of freedom $\bar{v}$ for 1D and 2D $\phi^4$-models, and for lattice sizes: $N=10\times 10$ (circles), $N=20\times 20$ (squares), $N=30\times 30$ (triangles) in the 2D case, and $N=900$ (rhombs) in the 1D case. The vertical dashed line indicates the phase transition point at $\bar{v}\simeq 2.2$.}
\label{var-diverg}
\end{figure}

\begin{figure}[h!]
 \centering
 \includegraphics[scale=0.25,keepaspectratio=true]{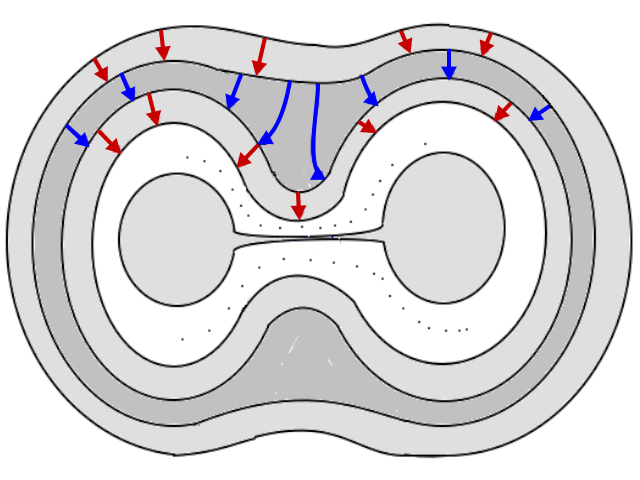}
\caption{ (Color online) Toy model representation of the possible geometrical origin of the peaks in Figs. \protect\ref{var-chi} and \protect\ref{var-diverg}. The first four lines pictorially represent "level sets" separated by the same potential energy interval. The first two external lines abstractly represent level sets $\Sigma_v$ at $v>v_c$ (above the phase transition). The third and fourth lines abstractly represent level sets $\Sigma_v$ at $v<v_c$ (below the phase transition) with a neck.  The variance of the length of the blue lines (corresponding to the formation of the neck) is larger than that of the red lines.}
\label{toymodel}
\end{figure}

Then, the two following quantities have been numerically computed along the mentioned  Monte Carlo Markov Chain:
 $\sigma^2(\chi)=\langle\chi^2\rangle_{\Sigma_v} - \langle\chi\rangle^2_{\Sigma_v}$ and $\sigma^2(\mathrm{div}\boldsymbol{\xi})=\langle(\mathrm{div}\boldsymbol{\xi})^2\rangle_{\Sigma_v} - \langle(\mathrm{div}\boldsymbol{\xi})\rangle^2_{\Sigma_v}$. These are functions of $N$ and of the specific potential energy ${\overline v}=V/N$.
The outcomes, reported in Figs. \ref{var-chi} and \ref{var-diverg}, show very different patterns in the $1D$ and $2D$ cases: monotonic  for the $1D$ case, non-monotonic displaying  cuspy points at ${\overline v}={\overline v}_c$ (the phase transition point) of $\sigma^2(\chi)$ and of $\sigma^2(\mathrm{div}\boldsymbol{\xi})$ for  the $2D$ case. As $\chi =1/\Vert\nabla V\Vert$ is locally proportional to the distance between nearby level sets, its variance is a measure of the total dishomogeneity of this distance, so that a peak of $\sigma^2(\chi)$ can be due to the formation of a "neck" in the $\{\Sigma_v \}_{v\in\mathbb{R}}$ foliation of configuration space. This is pictorially shown through the toy model of Fig.\ref{toymodel}. The same is true for $\sigma^2(\mathrm{div}\boldsymbol{\xi})$ since $\mathrm{div}\boldsymbol{\xi}$ is locally proportional to the variation of the area of a small surface element when a level set is transformed into a nearby one by the diffeomorphism in Eq.\eqref{hirschflow}.

\noindent{\bf Discussion. }  In spite of the absence of critical points of $V(q)$ of the $\phi^4$ model [Eq.\eqref{potfi4}] in correspondence with the phase transition potential energy density ${\overline v}_c$, we have here shown that this transition stems from an \textit{asymptotic} change of topology of both the $\Sigma_E$ and $\Sigma_v$. This leads to a more general formulation of the topological theory of phase transitions once a basic assumption of the theory is made explicit also in the $N\to\infty$ limit. This can be achieved by resorting to the explicit analytic representation \eqref{hirschflow} of the diffeomorphism $\boldsymbol{\xi}_N:\Sigma_{v,N}\subset\mathbb{R}^{N+1}\rightarrow\Sigma_{v^\prime,N}\subset\mathbb{R}^{N+1}$. Uniform convergence in $N$ of the sequence of  vector valued many-variable functions $\{\boldsymbol{\xi}_N\}_{N\in\mathbb{N}}$ can be used to define  \textit{asymptotic diffeomorphicity} in some class ${\cal{C}}^{l}$ of the  $\{\Sigma_v \}_{v\in\mathbb{R}}$ after the introduction of a suitable norm containing all the derivatives up to $(\partial^l\boldsymbol{\xi}_N/\partial x_{i_1}^{l_1}\dots\partial x_{i_k}^{l_k})$. Accordingly, in the theorems of Refs.\cite{prl1,TH1,TH2} the assumption of asymptotic diffeomorphicity of the $\{\Sigma_v \}_{v\in\mathbb{R}}$ has to be added to the hypothesis of diffeomorphicity just at any finite $N$. 
In this context it is worth mentioning that with a completely different approach also the phase transition of the $2D$ Ising model (which is of the same universality class of the $2D$ lattice $\phi^4$ model) is found to correspond to an asymptotic change of topology of suitable manifolds. This is found by proving that the analytic index of a given elliptic operator - acting among smooth sections of a vector bundle defined on a state manifold -  makes an integer jump at the transition temperature of the $2D$ Ising model \cite{rasetti1,rasetti2}. Hence the asymptotic change of topology of sections of the mentioned vector bundle stems from the Atiyah-Singer index theorem which states that the analytic index is equal to a topological index \cite{nakahara}. The extended versions of the theorems in  \cite{prl1,TH1,TH2} will be given elsewhere.
\hfill



\begin{thebibliography}{99}
\bibitem{nota} The parameters chosen here are the same of Ref.\protect\cite{CCP} where the critical energy density is shifted by a constant vaue of $10$.
\bibitem{nota3} The $N\to\infty$ extrapolation is safe because increasing $N$ essentially amounts to gluing togheter identical replicas of smaller lattices.
\bibitem{nota2} Notice that in this case transversality is absent, see \protect{\cite{hirsch}}.
\end{thebibliography}
\end{document}